\documentclass[12pt]{article}
\usepackage[utf8]{inputenc}

\usepackage{braket}
\usepackage{authblk}
\usepackage{amsmath}
\usepackage{cancel}
\usepackage{breqn}
\usepackage{xcolor}
\usepackage{stackengine}
\usepackage{hyperref}
\stackMath

\usepackage{feynmf}
\usepackage[numbers]{natbib}
 \usepackage[nottoc]{tocbibind}

\usepackage{feynmf}
\usepackage{multirow}
\usepackage{caption}
\usepackage{subcaption}

\def \ee{\end{equation}}
\def \be{\begin{equation}}
\def \eea{\end{eqnarray}}
\def \bea{\begin{eqnarray}}

\newcommand{\puc}{Instituto de F{\'i}sica, Pontificia Cat{\'o}lica Universidad de Chile, Av. Vicu{\~n}a Mackenna 4860, Santiago, Chile}
\newcommand{\vienaTU}{Institut f\"ur Theoretische Physik, TU Wien,
Wiedner Hauptstrasse 8-10, 1040 Vienna, Austria}

\newcommand{\SLU}{Research Centre for Theoretical Physics and Astrophysics, Institute of Physics, Silesian University in Opava,
Bezručovo náměstí 13, CZ-74601 Opava, Czech Republic.}
\newcommand{\UPV}{Instituto Universitario de Matemática Pura y Aplicada,
Universitat Polit\`ecnica de Val\`encia, Valencia 46022, Spain.}
\newcommand{\UIBK}{Universität Innsbruck, Fakultät für Mathematik, Informatik und Physik, Institut für Experimentalphysik, 6020 Innsbruck, Austria}

\title{Casimir Effect and Gravitational Balance:\\
a Search for Stable Configurations}

\author[1,2]{Leonardo Bellinato Giacomelli}
\affil[1]{\vienaTU}
\affil[2]{\UIBK}

\author[1,3]{Benjamin Koch 
\thanks{E-mail: \href{mailto:benjamin.koch@tuwien.ac.at}{\nolinkurl{benjamin.koch@tuwien.ac.at}}}}
\affil[3]{\puc}

\author[1]{Iva Lovrekovic}

\author[4,5]{Angel Rincon
\thanks{E-mail: \href{mailto:angel.rincon@physics.slu.cz}{\nolinkurl{angel.rincon@physics.slu.cz}}}
}
\affil[4]{\SLU}
\affil[5]{\UPV}

\date{}

\begin{document}

\maketitle
\begin{abstract}
In this study, we examine the role of the repulsive Casimir force in counteracting the gravitational contraction of a thin, spherically symmetric shell. Our primary focus is to explore the possibility of achieving a stable, balanced configuration within the theoretically reliable weak-field limit. To this end, we consider various types of Casimir forces, including those generated by massless scalar fields, massive scalar fields, electromagnetic fields, and temperature-dependent fields.
\end{abstract}

\newpage 
\tableofcontents
\newpage 
\section{Introduction}
\subsection{Gravitational balance}

Gravitational balance is a fascinating area of research within astrophysics and general relativity that explores the processes by which massive objects, such as stars, balance out with their gravitational force or collapse under it~\cite{Tolman:1930ona,Bekenstein:1971ej,Nowakowski:2001zw,Joshi:2011zm,Belvedere:2012uc,Pretel:2020xuo,Misner:1973prb}.
Research in this area continues to push the boundaries of our understanding of the universe, offering tantalizing prospects for breakthroughs in both the theoretical and observational realms. It pushes our understanding of the laws of nature to its boundaries.

Among the possible forces that might stabilize a gravitational system, one has considered electromagnetic forces, strong forces, the Pauli exclusion principle, and quantum vacuum energy.
The former forces are directly associated with the coupling strength of the standard model of particle physics. The latter forces are of totally different nature, because their existence is directly linked to genuine quantum effects.

Among these candidate forces, the 
quantum vacuum energy is certainly
the least explored and least understood. 
One reason for our limited knowledge of vacuum energy is certainly the difficulty in measuring it~\cite{Lamoreaux:2005gf}. Another reason is the fact that the role of quantum vacuum energy is largely unclear when it is coupled to gravity. 
This lies at the core of the infamous cosmological constant problem.
This is the large discrepancy between the theoretical predictions of vacuum energy density from quantum field theory and the observed value of this density from astronomical observations at cosmological scales, which differ by approximately 120 orders of magnitude~\cite{Weinberg:1988cp}.
Understanding the interplay between vacuum energy density and gravity might also play a non-trivial role in the evolution process of compact gravitational objects~\cite{Reyes:2023fde,Reyes:2023ags}. This is the region of high curvature that we do not explore here.

It is thus an interesting research objective to learn more about vacuum energy in general (see Subsection \ref{subsec_Casimir}) and in particular in the context of a gravitational collapse for the thin shells similar to Israel (see \ref{subsec_CasimirCollapse} and the following). Due to the specific nature of the theory, we expect our observations to appear at extreme scales.

\subsection{Casimir effect}
\label{subsec_Casimir}

Studies of the Casimir effect span multiple disciplines. It is important in quantum field theory, atomic and molecular physics, condensed matter physics, gravity and cosmology, mathematical physics, and others~\cite{Casimir:1948dh,Bordag:2001qi}. For instance, in Quantum Field Theory, bag models of hadrons, the Casimir modes of quarks, and gluons significantly contribute to the total nucleon energy. In Kaluza-Klein theory, the Casimir effect provides an effective mechanism for the spontaneous compactification of extra dimensions. In Condensed Matter Physics, the Casimir effect can be attractive or repulsive between tightly aligned material boundaries, strongly depending on the geometry of the boundaries, the temperature, and the mechanical and electrical properties of the boundary surface \cite{Bordag:2009zz}.

In the bag model of Quantum Chromodynamics (QCD), the boundary conditions on the spinor and gluon fields are related through equations of motion. Essentially, there is a Lagrangian model for the bag, along with a phenomenological Lagrangian description for the quantum behavior of the fields within the bag. 
Although the MIT bag model, since its discovery long ago, has been largely replaced by newer methods in QCD to study hadrons, it was a successful phenomenological description in its time~\cite{Chodos:1974je,Chodos:1974pn}. The precision of this effective phenomenological description led to numerous discoveries~\cite{Barnes:1982tx,Fraga:2012fs}.

Boyer demonstrated the importance of geometry in calculating the Casimir force~\cite{Boyer:1968uf}. He investigated the quantum electromagnetic zero-point energy (known as the Casimir energy) of a conducting spherical shell. By analogy with the parallel plate configuration, he expected the negative contribution from zero-point energy to induce a collapse of the sphere, balanced by the electrostatic energy of the configuration. To his surprise, the Casimir energy contribution was positive. He considered quantum electromagnetic modes within the sphere.
Further studies yielded interesting results on the dependence on geometry ~\cite{Bordag:2001qi}. Specifically, the energy contribution for a rectangular cavity can vary between positive and negative values, depending on the dimensions of the cavity. In one case, it may be positive, while in another it may be negative.

Numerous studies on the conducting sphere initiated by Boyer~\cite{Boyer:1968uf} and followed by analyses using Green’s function methods~\cite{Balian:1977qr,Milton:1978sf}. These include the calculation of the vacuum energy for (i) spinors in the bag model~\cite{Milton:1980ke,Milton:1983wy}, (ii) massive scalar enclosed in the system~\cite{Bordag:1996ma}, (iii) and for fermions~\cite{Elizalde:1997hx}. For the massless scalar field inside a sphere, the Casimir energy is given by
\begin{align}E_{Cas}=\frac{\hbar c}{R}\left(0.0044+\frac{1}{630\pi}\left[\frac{1}{s}+\left(\frac{\mu cR}{\hbar}\right)\right]\right)\label{casscalar},\end{align} 
where sphere is of a radius R in D=3 dimensions with Dirichlet boundary conditions \cite{Bordag:2009zz}, and $s$ is an unphysical parameter which cancels after taking into account the contribution from the outside of the sphere. In addition, experimental studies of this effect have intensified recently~\cite{Haghmoradi:2024ytw}. From here on, we are going to set $\hbar=1=c$.

In the gravitational case, the analysis of quantum modes within an analogous conducting sphere is more complex and has not been extensively considered. However, studies have been conducted on gravitons in the Casimir box, showing a linear dependence compared to the analysis for scalars. 
In the scenario with two infinite parallel planes separated by distance $d$, linearized gravity, similar to electromagnetism, can be represented as two free massless scalar fields—one with Neumann boundary conditions and one with Dirichlet boundary conditions. 
The calculated partition function can then be obtained also by studying a  single massless
scalar field on an interval of length $2d$ with periodic boundary conditions
~\cite{Alessio:2020lpk}.

In the following sections, we apply the Casimir force to systems with gravitational interactions. The Casimir force, in its original form, was derived for metals. However, metals are relatively rare in astrophysical contexts. It is therefore important to note that there is an analogous force for polarizable molecules, known as the Casimir-Polder force~\cite{Casimir:1947kzi}, and, as demonstrated by Lifshitz, a similar force applies to any dielectric material~\cite{Lifshitz:1956zz,Bordag:2001qi}. These forces exhibit the same dependence on geometrical properties, such as distance, as the idealized Casimir force. As a result, the following discussions are relevant to all types of materials, not just metals.

\subsection{Thin shells as toy model for gravitational collapse}
\label{subsec_CasimirCollapse}

In this paper, on phenomenological grounds, we consider a scenario in which a thin spherical shell with non-vanishing mass counteracts the repulsive Casimir force due to its gravity. The Casimir energy arises from the fluctuations of the dynamical field both inside and outside the shell.

If we adopt the line of thinking used in the MIT bag model to a scenario involving gravity, we can attempt to model the interplay between zero-point energy and gravitational effects. To utilize the result for the Casimir energy of dynamical field inside a perfectly conducting sphere, we can imagine a model similar to a bag, but with quantum fluctuations occurring on both sides of the boundary. The sphere is massive and is described in flat space, experiencing gravitational effects due to its mass.
In realistic physical scenarios, this description would be too simplistic. It would be necessary to carefully determine the energy of particles fluctuating within the system and the scale at which these fluctuations would be sufficient to counterbalance the mass.

A very instructive toy model in this direction was introduced by Brevik~\cite{Brevikk}, who included the contribution of Casimir energy to the gravitational system of a shell by hand. In his model, the shell contracts to a minimum radius, at which point it encounters a barrier at the speed of light. Without the Casimir force, this setup would lead to the collapse of the shell.

\subsection{Thin shells as toy model for fundamental sub-structure}
\label{subsec_CasimirSub}

In his early contributions, Casimir, building on previous ideas from Lorentz, suggested that the interior of fundamental particles might be described by a thin shell, which is stabilized against repulsive electrostatic forces by vacuum forces~\cite{Carazza:1986}. This type of subatomic model also motivated Boyer and others to explore the spherical Casimir effect.

As it turned out, the spherical Casimir force was repulsive, invalidating the Casimir-Lorentz model for charged particles such as electrons~\cite{Boyer:1968uf,Milton:1978sf}. However, this type of "inner structure model" continues to attract researchers' interest~\cite{Puthoff:2006fg}.

In this context, one might also ask: "Under what circumstances could the repulsive Casimir force stabilize neutral particles, such as neutrinos, against gravitational collapse?" Naturally, one could begin to address this question by using the Casimir-Lorentz model and exploring which types of interactions would lead to such stabilization.

Since the inner structure of these particles remains unknown, any such study should remain agnostic about the specific type of Casimir forces to which the shell is sensitive and focus instead on identifying which interactions result in stabilization and which do not.

\subsection{The research question}

As a toy model, motivated by both astrophysical situations as in Subsection \ref{subsec_CasimirCollapse} and small-scale structure models as in Subsection \ref{subsec_CasimirSub}, we use a thin spherical shell that experiences both gravitational and vacuum effects simultaneously. Brevik had already implemented a description of this system, as mentioned in the previous section~\cite{Brevikk}. In his work, the focus is on a regime of strong curvature and large velocities. However, in this regime, the Newtonian concept of simply summing forces to obtain a total force is likely not valid or is at least obscured by significant theoretical uncertainties.

Therefore, in our work, we will focus on a regime of low curvature and small velocities, where the Newtonian approach of adding the Casimir force to the gravitational equations of motion is certainly valid. Consequently, we concentrate on the following research question:

{\it ``Under which circumstances can Casimir forces counteract gravitational attraction and lead to stable configurations, even in the non-relativistic, low-curvature limit?''}

In this regime of non-relativistic dynamics, we consider three distinct scenarios:
\begin{itemize}
    \item[$\alpha$)] The shell collapses, with gravitational forces overwhelming the Casimir force.
    \item[$\beta$)] The shell expands from its initial radius due to a stronger Casimir force.
    \item[$\gamma$)] The shell oscillates around a stable radius.
\end{itemize}
Our primary focus is to explore whether the dynamics of the Casimir shell is always destructive (either $\alpha$ or $\beta$), or if there exists a region in the parameter space and initial conditions where the shell is stable ($\gamma$). We will classify different regimes and types of Casimir forces according to the scenarios ($\alpha, \beta, \gamma$) they produce for the spherical massive shell.

The interplay between the Casimir effect and the cosmological constant is an interesting related question which has been discussed in \cite{Mahajan:2006mw,Ai:2024taz,Padmanabhan:2006ag,Padmanabhan:2006cj,Koch:2022cta}. However, in this paper we do not consider this effect.

\subsection{Structure of the paper}

In the following section, we will define the setup of a thin self-gravitating shell~\ref{sec_setup}. Then, 
we will explore this system for different types of Casimir forces and categorize it into scenarios $\alpha$, $\beta$, $\gamma$.
In particular, we will consider a massless scalar field, a massive scalar field, and a temperature dependence for massless scalar and electromagnetic fields.

\section{Results}
In this section, we explore various avenues to determine whether they have the potential to provide an oscillating scenario $\gamma)$.
\subsection{The setup}
\label{sec_setup}

If we have a vacuum region surrounding any spherically symmetric matter distribution, and we describe this vacuum region with a metric, there exist coordinates in which the metric takes the standard Schwarzschild form, with a parameter $M_0$ equal to the enclosed interior mass. This holds true even when the vacuum region itself is embedded in a spherically symmetric distribution of matter, as verified in \cite{Kim:2016osp}. 
It is important to note that when considering specific scenarios, we must be careful in interpreting Birkhoff's theorem (see Appendix {\bf{A}} for a discussion). For a metric with the diagonal line element:
\begin{align}
ds^2 = -A(r)dt^2 + B(r)dr^2 + r^2d\Omega^2
\end{align}
with a priori arbitrary functions \( A(r) \) and \( B(r) \), we can consider a thin spherical shell by first considering a shell with mass density \( \rho(r) \) and finite thickness, which supports tangential pressure \( p(r) \). However, there is no radial pressure component. If the shell consists of this specific "fluid" with zero radial stress and non-zero compressive stress, it is possible to have a static system. If the radial stress did not vanish, it would be required that at the boundaries with interior and exterior vacua, the stress matches to zero. In this model, the components of the energy-momentum tensor within the shell are \( T_{\mu\nu} = \text{diag}(\rho(r) A, 0, p(r)r^2, p(r) r^2 \sin^2\theta) \). In this scenario, in the interior region, one obtains
\cite{Kim:2016osp}
\begin{align}
ds^2=-\frac{r_S-2(m_i +m_S)G_N}{r_S-2m_i G_N}\left(1-\frac{2m_i G_N}{r}\right)dt^2+\frac{dr^2}{1-\frac{2m_i G_N}{r}}+r^2 d\Omega^2.
\end{align}
The metric in the exterior region is 
\begin{align}
ds^2=-\left(1-\frac{2(m_i +m_S)G_N}{r}\right)dt^2+\frac{dr^2}{1-\frac{2(m_i +m_S)G_N}{r}}+r^2 d\Omega^2.
\end{align}
Here, \( d\Omega^2 = d\theta^2 + \sin^2\theta d\phi^2 \), \( m_i \) is the enclosed mass, \( m_S \) is the mass of the shell, and \( r_S \) is the radial position of the shell. This is the metric we will consider in our calculations, while setting the mass inside the shell \( m_i \) to zero ($m_{\text{i}} = 0$). When $m_i=0$, the spacetime inside the cavity is flat rather than Schwarzschild. When $m_i\neq 0$, 
one could perform a coordinate rescaling of "t" component (or equivalently set \(\frac{r_S - 2(m_i + m_S) G_N}{r_S - 2m_i G_N}\) to 1) to recover the Schwarzschild line element. However, this would be incorrect while leaving the line element outside the shell unchanged, as it would lead to an unphysical discontinuity in the time coordinate and in \( A(r) \) across the shell.
Now, the line elements read
\begin{align}
ds^2&=-\frac{r_S-2m_S G_N}{r_S}dt^2+dr^2+r^2 d\Omega^2,&& \text{  in the interior and, } \\
ds^2&=-\left(1-\frac{2m_S G_N}{r}\right)dt^2+\frac{dr^2}{1-\frac{2m_S G_N}{r}}+r^2 d\Omega^2
,&& \text{  in the exterior.}
\end{align}
\textcolor{black}{If we incorporate cosmological constant the interior and exterior metric read
\begin{align}\label{eq_ds2int}
ds^2&=-\frac{r_S-2m_S G_N}{r_S}(1-\frac{\Lambda}{3} r^2)dt^2+\frac{dr^2}{1-\frac{\Lambda}{3} r^2}+r^2 d\Omega^2,&& \text{  in the interior and, } \\
ds^2&=-\left(1-\frac{2m_S G_N}{r}-\frac{\Lambda}{3} r^2\right)dt^2+\frac{dr^2}{1-\frac{2m_S G_N}{r}-\frac{\Lambda}{3} r^2}+r^2 d\Omega^2
,&& \text{  in the exterior.}
\end{align}\label{eq_ds2ext}
}

Next, we outline the equations of motion (EOM) for the shell presented in \cite{2} with this metric and explore its implications on the existence of a minimal radius within which the shell is not able to collapse further. The metric enters the derivation of the EOM through the Christoffel symbols. The relevant Christoffel symbols are:
$\Gamma_{t t}^{r} = \frac{m_S G_N}{r^2}, 
    \Gamma_{r r}^{r}  = 0$.
    Following the prescription of Israel, we can obtain the normal interior and exterior acceleration. Similarly to the Israel situation, we have the Schwarschild manifold in the exterior, however in the interior, the time-time component of the metric has the prefactor $h(t)=\frac{r_S -2m_S G_N}{r_S}$. 
    To consider a variable radius of the shell we denote
    \begin{align}
    r_S=R(s).
    \end{align}
Here, $s$ is proper time measured along the streamlines~\cite{Israel:1966rt}. 
If we denote the $\xi$ as intrinsic coordinates of the smooth hypersurface $\Sigma$ in Riemannian manifold $V$, and $ds$ an infinitesimal displacement in $\Sigma$, we can define $u^{\alpha}=\frac{d\xi^{\alpha}}{ds}$  as streamilnes of the dust particles, following the notation of Israel \cite{Israel:1966rt}.

We will denote the derivatives with respect to $s$ with a dot. Using the velocities of the shell and normal vectors from the exterior region 
$u^a_+=\frac{dx^a_+}{ds}$, $n_a^+$, and from the interior    
$u^a_-=\frac{dx^a_-}{ds}$, $n_a^-$, we can obtain
\begin{align}&\text{the exterior:} && \boldsymbol{a}_+=n_{\alpha}\frac{\delta u^{\alpha}}{\delta s}|^+=n_a\frac{du_+^a}{d s}\boldsymbol{n}, \end{align} 
\begin{align}
&\text{and the interior:} && \boldsymbol{a}_-=n_{\alpha}\frac{\delta u^{\alpha}}{\delta s}|^-=n_a\frac{du_-^a}{d s}\boldsymbol{n} 
\end{align}
normal acceleration. This leads to the exterior normal acceleration equivalent to the standard Israel result.
\begin{equation}\label{a+}
    a_+ = \frac{\ddot{R} +  \frac{m_S G_N}{R^2}}{(1 + \dot{R}^2 - \frac{2  m_S G_N}{ R})^{1/2}}, 
\end{equation}
while for the interior acceleration we obtain 
\begin{align}
a_-=\frac{1}{\sqrt{1-\frac{2m_S G_N}{R}}}\left( \frac{\Ddot{R}}{\sqrt{\Dot{R}^2+1}} -\frac{m_S G_N\sqrt{1+\Dot{R}^2}}{R(2m_S G_N-R)}\right).
\end{align}

\textcolor{black}{Including cosmological constant the exterior acceleration reads
\begin{equation}\label{a+}
    a_+ = \frac{\ddot{R} +  \frac{m_S G_N}{R^2}-\frac{\Lambda}{3} R^2}{(1 + \dot{R}^2 - \frac{2  m_S G_N}{ R}-\frac{\Lambda}{3} R^2)^{1/2}}, 
\end{equation}
while the interior acceleration becomes
\begin{align}
a_-=\frac{1}{\sqrt{(1-\frac{2m_S G_N}{R})(1+\dot{R}^2-\frac{\Lambda}{3} R^2})}\left( \ddot{R} -\frac{R^3 \frac{\Lambda}{3}+ G_N m_S(1+\dot{R}^2-\frac{\Lambda}{3} R^2)}{R(2m_S G_N-R)}\right).
\end{align}
}
If we now consider the radial Israel equation \cite{Israel:1966rt}
\begin{align}
n_{\alpha}\frac{\delta u^{\alpha}}{\delta s}|^++n_{\alpha}\frac{\delta u^{\alpha}}{\delta s}|^-=0
\label{vacuumeq}\end{align}
 we can insert the values for the $\boldsymbol{a}_-$ and $\boldsymbol{a}_+$ to obtain the EOM for the spherical shell of dust which combines exterior and interior, $\boldsymbol{a}_++\boldsymbol{a}_-=0$. 
As we commented above, this configuration can only be stabilized if the stress energy tensor provides some tangential pressure. In our considerations, the role of stabilization will be taken by 
the Casimir force in the radial direction, i.e. the vacuum fluctuations. 
It is possible to derive the equation (\ref{vacuumeq}) from Einstein's equations of motion; however, we have chosen to follow Israel's procedure.

\subsection{Massless scalar field}\label{Massless scalar field}

We wish to incorporate the effect of
quantum vacuum fluctuations on the motion of the shell.
In the low curvature non-relativistic regime this can be achieved by adding the corresponding Casimir force to the Israel's radial equation of motion of a shell. \textcolor{black}{(Furthermore, in three dimensions, the Casimir energy will be finite, as it is an odd number of dimensions, \cite{Bordag:2009zz}.)} Casimir energy in three dimensions, since it is odd number of dimensions will be finite, \cite{Bordag:2009zz}.

First, let's define the Casimir pressure $P$
as  $F$ divided by the surface area $A=4 \pi R^2$
\be
 P \equiv  \frac{F}{A}.
\ee
The Force, in turn, is given from the radial change in the Casimir energy $E$
\begin{align}
\label{pcasimir}
   P= -\frac{1}{4 \pi R^2} \frac{\partial E}{\partial R}.
\end{align}
The Casimir energy $E$ is the sum of Casimir energies $E_{\text{in}}$ and $E_{\text{out}}$ for the inside and the outside of the shell.
\textcolor{black}{
In addition, the Casimir energy of a massless scalar field over the whole space (inside the sphere and outside) is given by 
\begin{equation}
\label{ecasimir}  
E=\frac{C}{2R} .  
\end{equation}
}
We are going to denote the contribution of Casimir pressure with 
\begin{equation} \label{pres}
    P =  
    \frac{ C  }{8 \pi R^4},
\end{equation}
where $C$ is a  numerical constant.

To implement Casimir energy, we assume that:
\\
(i) Backreactions are sub-leading contributions.
This means that on the one hand, the space-time curvature is dominated by the mass of the shell, recognizing that Casimir energies typically much smaller than the energy corresponding to the rest mass. Thus, the local Casimir energy does not contribute to the total energy-momentum tensor appearing in Einstein's field equations.
On the other hand one also treats the background metric that is used to calculate the Casimir energy as approximately flat. 
In any realistic scenario, both approximations are perfectly reasonable.
\textcolor{black}{It is interesting to explore the potential impact of incorporating a cosmological constant, $\Lambda$, into the underlying solution. 
This can be done in two regimes:
\begin{itemize}
    \item Outer region:\\
    The influence of $\Lambda_e$ depends on whether we consider regions far from or near the origin. In the far-field case, the significance becomes evident when we recognize the existence of the cosmological horizon, such as the Hubble radius, which acts as a natural boundary for an observer’s accessible Universe. This boundary may, in principle, modify vacuum energy modes in a manner analogous to the Casimir effect \cite{2025Symm...17..634G}.
    Still, the numerical value of the observed cosmological constant $\Lambda_e$ is so small that it does not play a role at non-cosmological distances. 
    \item Inner region:\\
One might think that the vacuum energy produced by quantum fluctuations can mimic a cosmological constant in the interior region $\Lambda_i$. If this were the case, one should include this $\Lambda_i$ into the calculations, for self consistency. However, the stress--energy tensor induced by vacuum fluctuations  in a static spherically symmetric setting $T_{\mu \nu}$~\cite{Bordag:2001qi}, is completely different from the tensor structure of an interior cosmological constant $\propto \Lambda_i g_{\mu \nu}$. Thus, for some parts of the calculations we will include the possibility of such an interior constant $\Lambda_i$, but we emphasize that it's existence can not be justified by fluctuations of the quantum vacuum obeying static spherically symmetric boundary conditions.
\end{itemize}
In any case, although we expect minimal change, we have included the cosmological constant term to account for such a modification. 
}
\\
(ii) In the regime of the above assumptions, the Casimir force is introduced as an additional term in the equations of motion, 
rendering the treatment non-relativistic. \\
Under these assumptions, 
the gravitational collapse of a shell in the presence of the Casimir force is described by~\cite{1986NCimB..94...54V}
\begin{align} \label{EOM1}
    M \Bigg[\frac{1}{2}\Bigl(a_+ + a_- \Bigl) 
    \Bigg] = (4\pi R^2) P,
\end{align}
where $m_S=M$ is the total proper mass of the shell, the term in square brackets represents the mean acceleration and the right-hand side of the last equation represents the force.
The above equation (\ref{EOM1}) is obtained by requiring the equilibrium of forces in the system, within the classical regime. On the left-hand side of the equation, we have classical effects, while on the right-hand side, we have macroscopic effects arising from the Casimir force, $P = P(C)$, which is a quantum effect.

To ensure the validity of this limit, we are considering a low curvature regime.

%
Note also that the last equation is reduced to the simplest case by neglecting the Casimir force. This approach, for the scenario when the metric in the interior is $diag(-1,1,1,1)$, can be found in \cite{Brevikk}.
Now, combining \eqref{pres}, \eqref{EOM1} and $M$, we can write the EOM as
\begin{align}
    \frac{1}{2}\sigma \left(a_+ + a_-\right) = \frac{C}{8 \pi R^4},
\end{align}
where $\sigma$ is the mass density per unit area (i.e., $\sigma\equiv M/A$).
Multiplying by $2/\sigma$ and writing the accelerations explicitly one gets
\begin{align}\label{eq_MasterM0}
\frac{m_S G_N \sqrt{1-\frac{2 m_S G_N}{R}} \sqrt{\Dot{R}^2+1}}{(R-2 m_S G_N)^2}+\frac{\Ddot{R}}{\sqrt{1-\frac{2 m_S G_N}{R}} \sqrt{\Dot{R}^2+1}}+\frac{\frac{m_S G_N}{R^2}+\Ddot{R}}{\sqrt{-\frac{2 m_S G_N}{R}+\Dot{R}^2+1}}=\frac{C}{m_S R^{2}}.
\end{align}
\textcolor{black}{After inclusion of cosmological constant this becomes 
\begin{align}& \frac{\ddot{R} +  \frac{m_S G_N}{R^2}-\frac{\Lambda_e}{3} R^2}{(1 + \dot{R}^2 - \frac{2  m_S G_N}{ R}-\frac{\Lambda_e}{3} R^2)^{1/2}}+
\\+&
\frac{1}{\sqrt{(1-\frac{2m_S G_N}{R})(1+\dot{R}^2-\frac{\Lambda_i}{3} R^2})}\left( \ddot{R} -\frac{R^3 \frac{\Lambda_i}{3}+ G_N m_S(1+\dot{R}^2-\frac{\Lambda_i}{3} R^2)}{R(2m_S G_N-R)}\right)=\frac{C}{m_S R^2}.
\nonumber\end{align}
Here we have named the cosmological constant in the interior as $\Lambda_i$ and cosmological constant in the exterior as $\Lambda_e$. If we want to include the cosmological constant to the regions where the vacuum fluctuations contribute with Casimir effect, we have to include it in the interior and the exterior. Distinguishing them allows us to switch them off independently. 
}

The Casimir force ordinarily includes its two-loop radiative corrections. However, these corrections do not account for significant differences in the dynamics of the system and will thus be neglected in further analysis. \\
\\
We wish to investigate the existence of a minimal radius resulting from the interplay between the inward-pointing gravitational force and the outward-pointing Casimir force. To do so, we make the simplifying assumption that the initial velocity of the shell is $\dot{R} = 0$, an
\begin{align}\label{eq_MasterMassless}
    \left[
    \frac{m_S G_N}{R^2} \left(  1 + \frac{1}{\Bigl(  1 - \frac{2 m_S G_N}{R} \Bigl) }  \right) + 2 \ddot{R}
    \right] 
    \frac{1}{\sqrt{ 1 - \frac{2 m_S G_N}{R}}} 
    &= \frac{C}{ m_S R^2}.
\end{align}
\begin{footnotesize}
\textcolor{black}{Including cosmological constant this reads \begin{align}\frac{\ddot{R} +  \frac{m_S G_N}{R^2}-\frac{\Lambda_e}{3} R^2}{(1  - \frac{2  m_S G_N}{ R}-\frac{\Lambda_e}{3} R^2)^{1/2}}+
\frac{1}{\sqrt{(1-\frac{2m_S G_N}{R})(1-\frac{\Lambda_i}{3} R^2})}\left( \ddot{R} -\frac{R^3 \frac{\Lambda_i}{3}+ G_N m_S(1-\frac{\Lambda_i}{3} R^2)}{R(2m_S G_N-R)}\right)=\frac{C}{m_S R^2}.\label{eq26}
\end{align}
}
\end{footnotesize}
Further, we are only interested in dynamics well outside the Schwarzschild radius $R>2m_S G_N$, where both the Newtonian and non-relativistic approximations can be trusted, and neglecting backreactions is well justified.

In this limit, we can expand the left-hand side of (\ref{eq_MasterMassless}) in inverse powers of $R$. This yields the equation for slow motions in the far region:
\be\label{eq_ddrml}
\ddot R= \frac{3 G_N m_S(C+ 2 G_N m_S^2)-2(C- 2 G_N m_S^2) R}{2 m_S R^4}.
\ee
The ``far region'' is defined by
\be\label{eq_Far}
R>> 2 m_S G_N.
\ee
Now, we seek for a stability condition.
One necessary condition for a stable configuration $\gamma)$ is the existence of a ``rest'' radius $R_r$ which has vanishing acceleration
\be\label{eq_rdd0}
\ddot R|_{R=R_r}=0.
\ee
For the case of (\ref{eq_ddrml}),
this condition is solved by
\bea\label{eq_rrsol}
R_r=\frac{G_N m_S(C+2 G_N m_S^2)}{C-2 G_N m_S^2}.
\eea
The second condition for stability, is
that the dynamics is attractive towards the rest radius, in some finite vicinity of $R_r$. This condition implies
\be\label{eq_dddrdr}
\left. \frac{d(\ddot R)}{dR}\right|_{R_r}<0.
\ee
Using (\ref{eq_rrsol}) to evaluate the left hand side of (\ref{eq_dddrdr}) we find
\be\label{eq_dddrdrm0}
\left. \frac{d(\ddot R)}{dR}\right|_{R_r}= \frac{(C-2 G_N m_S^2)^4}{2 G_N^3 m_S^4 (C+2 G_N m_S^2)^3}>0.
\ee
We remember that $C>0$.
Thus, (\ref{eq_dddrdrm0}) is in contradiction with (\ref{eq_dddrdr}), which means that a stable balance between the Casimir force and the gravitational dynamics is not possible in the Newtonian regime of a massless scalar particle (\ref{eq_MasterM0}).
A numerical analysis shows that the same is true if one explores slow motion for all radii $R> 2 m_S G_N$.
This can nicely be exemplified with figure \ref{fig_rddofr}.
\begin{figure}[h]
    \centering
    \includegraphics[width=14cm]{Rddm02.png}        
    \caption{Dimensionless acceleration $\ddot R/m_p$ as a function of dimensionless radius $R m_p$ for (\ref{eq_MasterM0}). The parameter $C$ is chosen to be $0.0028$.
    The blue and orange curves are for values of
     $m_S=0.034\, m_p,$and $ m_S=0.04 \,m_p$ respectively.
     }
    \label{fig_rddofr}
\end{figure}
Since in the true Casimir effect $C<0.1$,
all scenarios where the shell is heavier than the Planck mass are represented by the green curve. Light shells instead $m_S^2<C/(2G_N)$,
would be represented by the blue or orange cuver in figure \ref{fig_rddofr}.
 In any case,
the stability condition (\ref{eq_dddrdr}) is never fulfilled in the massless case. This means that in this case it is not possible to realize scenario $\gamma)$ (only $\alpha)$ or $\beta)$). 

\textcolor{black}{For completeness, the effects of the cosmological constant from (\ref{eq26}) are given in the appendix C.}
\newpage

\subsection{Massive scalar field}

\begin{figure}[h]
    \centering
    \includegraphics[width=14cm]{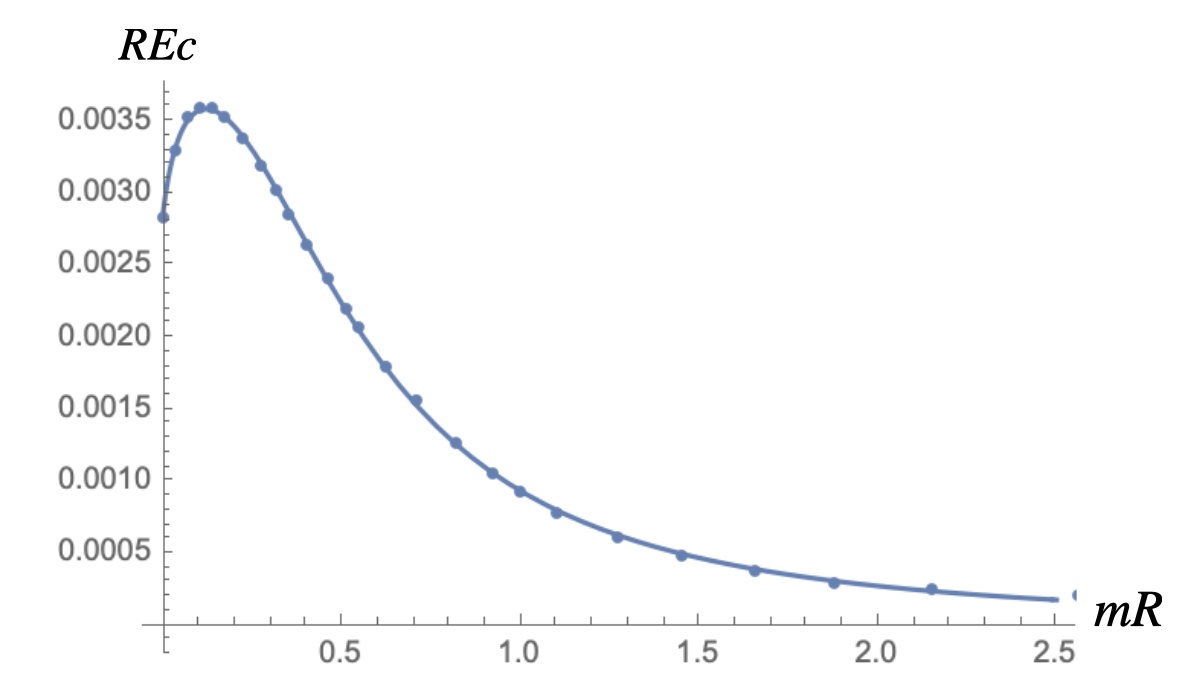}        
    \caption{ 
    Casimir force in a spherical shell produced by a massive scalar field.
    The points are the numerical results given in \cite{Bordag:2001qi}.
    The continuous curve is the fit given in equation (\ref{eq_fitREm}).
    \label{fig_rEforM}}
\end{figure}

In the previous section, we explored the influence of the Casimir force of a massless scalar field on the dynamics of the gravitational collapse of a shell. The analysis shows that there exists no stable minimal radius. 
We now expand the EOM to include the Casimir force for a massive scalar field.
 The force is pointed outward and the corresponding energy can be calculated numerically (see page 100 of~\cite{Bordag:2001qi}.
This result is well fitted by the phenomenological function of the dimensionless product of radius $R$ and energy $E_m$ as a function of the dimensionless product of mass $m_\phi$ times radius $R$
\be\label{eq_fitREm}
R E_m=\frac{a+ b \,(R m_\phi)}{1+ c\, (R m_\phi)+ d (R m_\phi)^3}.
\ee
Where the resulting dimensionless fitting parameters are $a=0.0029,\, b=0.0674, \, c=16.28,\, d=57.55$.
This fit is shown in figure \ref{fig_rEforM}.
Equipped with this relation, we can proceed to calculate the Casimir force via
\be
F_{m_\phi}=-\frac{d E_m}{dR}.
\ee
This force is then divided by the mass of the shell $m_S$ and inserted on the RHS of the relations (\ref{eq_MasterM0} and \ref{eq_MasterMassless}).
This gives the equation of motion for the shell with non-relativistic velocities
\bea\label{eq_ddRm}
\ddot R&=& \frac{a+m_\phi R(2 a c + m_\phi R(bc + d m_\phi R(4 a + 3 b m_\phi R)))}{2 m_S R^2 (1+c m_\phi R+d m_\phi^3 R^3)^{2}}\sqrt{1- \frac{2 m_S G_N}{R}} \\ \nonumber
&&-\frac{G_N m_S}{2 R^2}-\frac{G_N m_S}{2R^2(1-\frac{2 G_N m_S}{R})}.
\eea
The radii with vanishing acceleration are found by imposing (\ref{eq_rdd0}).
This can be solved analytically in the far region, yielding
\be\label{eq_Rsfarm}
r_S= \frac{\sqrt{6 bd G_N m_S^2 m_\phi^2-7 d^2 G_N^4 m_S^6 m_\phi^4}-d G_N^2 m_S^3 m_\phi^2}{2 d G_N m_S^2 m_\phi^2}.
\ee
For consistency, we have to demand that this radius is both real and much larger than the Schwarzschild radius of the mass $m_S$.
Interestingly, this implies an upper limit for the mass of the shell as a function of $m_\phi$
\be\label{eq_msMax}
m_S^4<\frac{3b}{16 d G_N^3 m_\phi^2}. 
\ee
Next we have to check whether the point of vanishing acceleration is actually stable.
This condition can be implemented in terms of the relation (\ref{eq_dddrdr}).
In this region we find
\be\label{eq_ddRform}
\left.\frac{d \ddot R}{d R}\right|_{r_S}=
16 d^4 G_N^5 m_S^9 m_\phi^8\frac{G_N m_S \sqrt{d G_N m_S^2 m_\phi^2(6 b - 7 d G_N^3 m_S^4 m_\phi^2)}+ 7 d G_N^3 m_S^4 m_\phi^2-6b}{(\sqrt{d G_N m_S^2 m_\phi^2(6b-7dG_N^3 m_S^4 m_\phi^2)}-dG_N^2 m_S^3 m_\phi^2)^5}.
\ee
The right hand side of this relation is negative if
\be\label{eq_msMax3}
m_S^4 < \frac{6 b}{7 d G_N^3 m_\phi^2},
\ee
which is already assured by imposing (\ref{eq_msMax}).
Thus, from the analytic approximations (\ref{eq_Rsfarm}-\ref{eq_ddRform}) we conclude that the Casimir force for massive fields could produce a meta-stable configuration (in the sense of dependence on the initial conditions) in the interplay with the gravitational attraction, if the condition
(\ref{eq_msMax}) is satisfied.

Next, we proceed to check this analytical intuition numerically by plotting (\ref{eq_ddRm}) as a function of $R m_p$. The result is shown in Figure \ref{fig_Rddm}.
From (\ref{eq_msMax}) and (\ref{eq_msMax3}) we see that the finite value of $b$ is crucial for the existence of a stable configuration. 
Since this fitting parameter also enters the energy condition, (e.g. the Null Energy Condition (NEC)) one can check for correlations between NEC violations and the existence of stable configurations.
We explore this scenario in Appendix D.
\begin{figure}[h]
    \centering
    \includegraphics[width=14cm]{Rddm22.png}        
    \caption{   
    Dimensionless acceleration $\ddot R/m_p$ (\ref{eq_ddRm}) as a function of $R m_p$ for $m_S=0.034\, m_p$.
    The blue curve is for $m_\phi=2.9 \,m_p$. The green curve is for $m_\phi=3.455\, m_p$. The orange curve is for $m_\phi=3.7 \,m_p$.
    The zeros of these curves are the solutions of $\ddot R=0$. Only the blue curve has a meta-stable configuration.
Interestingly, this meta-stable configuration establishes approximately at a radius which is close to the Compton wavelength of the scalar $\lambda\approx 1/m_\phi$, which is for any known particle much smaller than the physical radius of a macroscopic or micro meter sphere.
    \label{fig_Rddm}}
\end{figure}

To get an idea of which type of shells could be stabilized, let's assume that the stabilizing scalar field is the Higgs field with $m_\phi = 125~$GeV. The Casimir force for this field could stabilize, under the right initial conditions, a spherical shell with a mass smaller than $m_S < 10^{25}~$GeV $\approx 0.2~$kg. However, the stable radius for this configuration would be eight orders of magnitude smaller than the proton radius. This is extremely small, even though it is still three orders of magnitude larger than the Schwarzschild radius of this configuration.

This result can be seen from two perspectives. On the one hand, 
it is encouraging because it could be interpreted as a hint of a mechanism stabilizing the gravity-matter interplay at very high densities. On the other hand, it applies classical mechanics at distance scales that are deeply quantum. Thus, even though we found a scenario that realizes the stable scenario $\gamma)$, it has to be taken with a big grain of salt.

Note that another candidate for scalar fields could be cosmological fields such as the inflaton. These could provide very different masses and radii. However, since we do not know how to implement boundary conditions for this field, we do not explore this possibility further at this point.

\subsection{Temperature dependence}

In this section, we consider if there is a balance of the thin spherical shell and the repulsive Casimir force due to the fluctuation of massless scalar fields in the sphere at non-vanishing temperature. To model this behavior, we are going to use the acceleration from the interior and the exterior of the shell as above, while on the right-hand side, we implement the contribution from the Casimir force at finite temperature. The Casimir energy of the thin spherical shell in the limit of low and high temperature has been studied, and a review can be found in \cite{Bordag:2009zz}.
\textcolor{black}{
At this point, becomes relevant to reinforce why we consider the study of massless or massive scalar field, with non-vanishing temperature.
Naively, we have  neglecting excitations. In practice, the field includes real particles, usually in thermal equilibrium. This demands a statistical ensemble at temperature $T$ with a probability distribution. The ensemble's energy in the presence of boundaries defines the Casimir energy at finite temperature.
Thus, the finite-temperature scenario is physically more interesting and realistic, as it allows us to properly define the corresponding thermodynamics (first by introducing the partition function $Z$) and then deriving the remaining thermodynamic properties, such as the free energy, pressure, and entropy, to name a few.
}
\\

\subsubsection{High temperature \textcolor{black}{limit} for scalar field}

In the high temperature limit, the EOM for the shell is given by
\begin{align}\label{eq_eomHighT1}
    &\frac{\Ddot{R} +  \frac{m_S G_N}{R^2}}{(1 + \Dot{R}^2 - \frac{2  m_S G_N}{ R})^{1/2}}+\frac{1}{\sqrt{1-\frac{2m_S G_N}{R}}}\left( \frac{\Ddot{R}}{\sqrt{\Dot{R}^2+1}} -\frac{m_S G_N\sqrt{1+\Dot{R}^2}}{R(2m_S G_N-R)}\right) =  -\frac{1}{m_S} \frac{T}{24 R} 
\end{align} 
Like before, we are interested in the slow motion regime, for which we impose $\dot{R} = 0$. Furthermore, we remember that the entire classical approach without backreaction is only reliable in the Newtonian regime $R \gg m_S G_N$. In this expansion, we get from (\ref{eq_eomHighT1})
\be
\ddot R=-\frac{1}{m_S} \frac{T}{24 R} -G_N\frac{m_S}{R^2}.
\ee
From this expansion, it becomes clear that in the high-temperature limit, both the gravitational and the Casimir interactions contribute to an inwards-directed pull. Thus, in this regime, no stable configuration is possible; instead, only scenario $\alpha$) is realized.

\subsubsection{Low temperature \textcolor{black}{limit} for scalar field}

The Casimir energy for a scalar field in the low temperature \textcolor{black}{limit} is 
\textcolor{black}{
(see section 9.5 of ~\cite{Bordag:2009zz}, section "Spherical shell at nonzero temperature") 
}
$$
\Delta_T\mathcal{F}^D=\frac{\pi^3R}{6}(k_BT)^2-\zeta_R(3)R^2(k_BT)^3.
$$ 
 Here, the index $D$ denotes Dirichlet boundary conditions imposed on the sphere. The Boltzman constant is denoted with
 $k_B$  and $\zeta_R(3)$ is the Riemman zeta function. Combining the energy at zero temperature and the thermal correction, we obtain the total contribution from Casimir energy at low temperature. That gives the corresponding radial force acting on the spherical shell, which we insert on the right-hand side of the equation. This leads to
\begin{align}
     \frac{\Ddot{R} +  \frac{m_S G_N}{R^2}}{(1 + \Dot{R}^2 - \frac{2  m_S G_N}{ R})^{1/2}}+&\frac{1}{\sqrt{1-\frac{2m_S G_N}{R}}}\left( \frac{\Ddot{R}}{\sqrt{\Dot{R}^2+1}} -\frac{m_S G_N\sqrt{1+\Dot{R}^2}}{R(2m_S G_N-R)}\right) 
     \nonumber
     \\ 
     &=  \frac{C}{m_S  R^2}-\frac{ \frac{1}{6} \pi ^3  T^2-2R T^3 \zeta(3) }{m_S} 
\end{align}
 In the limit $T \rightarrow 0$, the EOM reduces to the case of a temperature independent massless scalar field. 

\begin{figure}[tbph]
\centering
\includegraphics[width=0.76\linewidth]{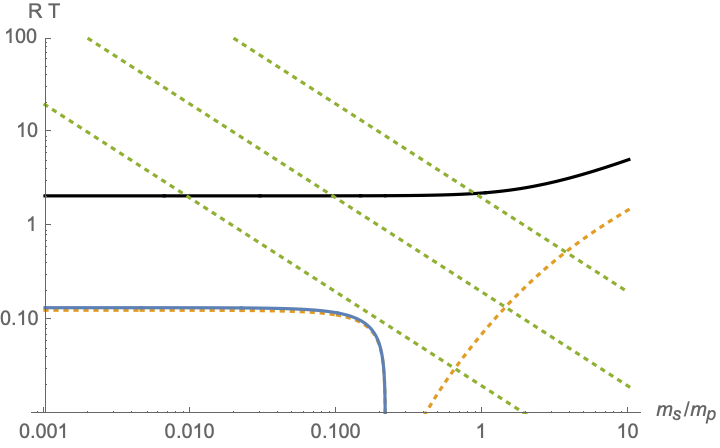}  %
\caption{ Value of the roots $T R_i$
as a function of $m_S/m_p$, where $m_p=1/\sqrt{G_N}$. The black line is the solution $R_1$, the blue line is the stable solution $R_3$. The solution $R_2$
is only plotted as dotted line, because for $m_S^2<G_N$ it is negative and for $m_S^2>G_N$ it is imaginary.
The green lines are $1/(m_S 200 ), \, 1/(m_S 20), 1/(m_S2)$.  They represent the underlying criterium of large radius expansion, namely that we only should trust scenarios below the green line which means temperatures below the Planck energy.\label{fig_RTofmsovermp}}
\end{figure}

As in the previous cases we investigate the conditions for which $\Ddot{R}  = 0 = \dot{R}$. The stable $R>0$ solutions for which this conditions are satisfied would be the sought after minimal radii $R_{\text{min}}$. Again, we first impose $\dot{R} = 0$ and solve for $\Ddot{R}$
\bea\nonumber
    \Ddot{R} &=& \frac{1}{2} \sqrt{1-\frac{2 m_S G_N}{R}} \left[\frac{ \frac{C}{R^2}-\left(\frac{1}{6} \pi ^3 T^2-2R T^3 \zeta(3)\right)}{m_S}-\frac{m_S G_N}{R \sqrt{1-\frac{2 m_S G_N}{R}}}\left( \frac{1}{R}-\frac{1}{2 m_S G_N-R}\right)\right] \\ 
    &\approx &
    \frac{T^2}{m_S  R^2}\left[    \frac{C-G_N m_S^2}{T^2 }- \frac{\pi^3 R^2}{12}+ \frac{R^3 T \zeta(3)}{1}\right]\, .\label{eq_scalarLowTemp}
\eea
Before solving the right hand side for $\Ddot{R} = 0$, let's discuss the apparent properties of these three roots. 
Since the leading term in the bracket $\sim R^3$ is positive, the largest positive root will always be unstable. If the $\sim cte.$ term
is also positive, there can be two positive roots, from which the smaller one would be stable.
If the $\sim cte.$ term is negative,
there could be one or three real roots. In the former case, the root is unstable. The latter case does not happen because if at some radius the $\sim R^2$ term dominates, it does so for positive and negative $R$, in particular since there is no $\sim R$ term.
The analytic expressions for the roots of the above equation (\ref{eq_scalarLowTemp}) we present in (\ref{eq_R0forTscalar}), in the Appendix.
The number of real roots can be shown to only depend on the parameter $m_S$. 
In Figure \ref{fig_RTofmsovermp} we plot ($T\cdot $(\ref{eq_R0forTscalar})) as a function of the dimensionless ratio \textcolor{black}{$m_s/m_p$}.
Figure \ref{fig_RTofmsovermp} shows nicely that a stable solution is only possible for light shells, as long as $C>G_N m_S^2$.
To illustrate this effect, we plot $\ddot R$
as a function of $R$, for different temperatures.

\begin{figure}[tbph]
\centering
\includegraphics[width=0.78\linewidth]{RddvonT.png} %
\caption{Dimensionless acceleration $\ddot R/m_p$ as a function of the dimensionless radius $R m_p$ as given from equation (\ref{eq_scalarLowTemp}).
Different colors correspond to different temperatures: blue ($T=0.01\, m_p$,), green ($T=0.1\, m_p$), and orange ($T=0.15\, m_p$).
   \label{temp1}
}
\end{figure}

From Fig.\eqref{temp1} we confirm that a stable configuration can be found for small $m_S$, even when no large radius expansion is imposed.
The stability can be seen from the fact that the radial acceleration is negative for $R > R_{\text{min}}$ and positive for $R < R_{\text{min}}$. Therefore, the interplay between the temperature dependent Casimir force of a massless scalar field with the gravitational attraction caused by the massive shell, can stabilize the system,
which corresponds to scenario $\gamma)$.

\subsubsection{Electromagnetic Casimir at low temperature and large $R\gg r_S$}

Since the scalar Casimir effect at low temperatures yielded the desirable scenario $\gamma$), it is reasonable to hope for a similar result when electromagnetic interactions are considered. The low-temperature corrections to the electromagnetic Casimir effect are, however, simply
 $\sim T^4$~\cite{Bordag:2009zz}. Thus, the motion of the shell is described by
\be
\ddot R= \frac{C}{m_S R^2}-G_N \frac{m_S}{R^2}+\frac{1}{5}\pi^3 R^2 (k_B T)^4.
\ee
The radius with zero acceleration is
\be
R_0^2=\frac{\sqrt{5}\sqrt{G_N m_S^2 - C}}{k_B^2 \sqrt{m_S}\pi^{3/2}T^2}.
\ee
The stability criterium at this radius is
\be
\left.\frac{d \ddot R}{d R}\right|_{R_0}=4 k_B^3\frac{(G_N m_S-C/m_S)^{1/4} \pi^{9/4}T^3}{5^{3/4}}>0.
\ee
Thus, if $m_S^2<C/G_N$ \textcolor{black}{($m_S^2/m_P^2 < C$)}, the configuration always explodes, which corresponds to scenario $\beta)$.
Instead, if $m_S^2>C/G_N$ \textcolor{black}{($m_S^2/m_P^2 > C$)}, the configuration is unstable in the sense that it depends on the initial condition, whether the outcome is the scenario $\alpha)$ or $\beta)$. A stable scenario $\gamma)$ is not possible here.


\section{Conclusion}

In the present work, we have investigated the dynamics of the gravitational collapse of a shell in the presence of the Casimir force in four dimensions considering different types of vacuum fluctuations:

\begin{itemize}
    \item[i)] massless scalar fields,
    \item[ii)] massive scalar fields,
    \item[iii)] massless scalar fields at high temperature limit,
    \item[iv)] massless scalar fields at low temperature limit,
    \item[v)] electromagnetic fields at low temperature limit.
\end{itemize}
Specifically, we have investigated, among other things, the existence of a minimal radius as a result of the interplay between
the inward-pointing gravitational force and the outward-pointing Casimir force. 
We defined the minimal radius as the point at which the acceleration and velocity of the collapsing shell can reach $\dot{R} = 0 = \Ddot{R}$. To evaluate the stability of the minimal radius $R_{\text{min}}$, we explored the derivatives of the acceleration $\Ddot{R}$ with respect to the radius $R$ at this point of interest.
Finally, we have illustrated our results with some figures.

In two out of the five scenarios that we analyzed we obtain the stable region, while in the other three scenarios no stability region is found: 
(i) when the Casimir force is produced by the massless scalar particles at the zero temperature, there is no radius at which the thin spherical shell would oscillate around a minimum. (ii) When the Casimir force is caused by massive scalar particles, the stable region is obtained when the mass shell is lighter than the fluctuating massive particle. (iii) For the thin spherical shell of the mass $m_S$ with the massless scalars at the low non-zero temperature, we obtain a stable region. The region satisfies the conditions that the radius of the shell is bigger than the two masses of the shell, $R>2m_S$, and that low-temperature behavior for the thermal corrections can be used $k_B TR \ll 1$ \cite{Bordag:2009zz}. (iv) In the limit when the Casimir energy is expanded around a high-temperature limit, we do not obtain a stable region. The same negative result is true for
the electromagnetic Casimir effect in the low temperature limit~(v).

Note that due to the large theoretical uncertainties on the interplay between quantum effects and general relativity we chose to study 
 this simple model in the regime where we can expect that all assumptions still hold, namely in the regime of weak gravity and non-relativistic motion. In this regime, we do not find realistic scenarios on our model, where the Casimir force could stabilize a shell against the gravitational pull.

\section{Acknowledgements}

The work of IL was supported by the Hertha Firnberg grant T1269-N and Elise Richter grant V 1052-N by the Austrian Science Fund, FWF.
A. R. acknowledges financial support from the Silesian University in Opava. 
The creation of this article was supported by the grant program Vouchers for Universities in the Moravian-Silesian Region (registration number CZ.10.03.01/00/23\_042/0000390).

\section*{Appendix}

\subsection*{A $\quad$ Remark on Birkhoff's theorem}
\label{subsecBirk}

According to Birkhoff's theorem \cite{1923rmp..book.....B}, in Einstein's theory of relativity, when we have a symmetric mass distribution that is not rotating in empty space we can describe the gravitational field outside this distribution using the Schwarzschild metric. In essence, this metric enables us to simplify the understanding and calculation of effects in several situations and has important implications. Birkhoff's theorem can be seen as a characterization of the structure of vacuum spacetimes that exhibit local spherical symmetry. Its significance lies in the fact that it serves not only as a classical cornerstone of general relativity but also as a vital tool in the fields of gravitational physics and cosmology.
In particular, Birkhoff's theorem is conventionally employed for the following purposes:
i) to establish that the empty spacetime within a spherically symmetric mass shell corresponds to Minkowski spacetime,
ii) to demonstrate that spherical gravitational collapse does not emit gravitational waves,
iii) to finalize the proof of the classical uniqueness theorem for static black holes \cite{Israel:1967wq}.
Given its significance, it becomes evident that Birkhoff's theorem plays a crucial role in the context of spherically symmetric spacetimes within the framework of general relativity.

There is some discussion in the literature concerning the applicability an validity of Birkhoffs' theorem.
In particular, Israel \cite{2} and Brevik \cite{Brevikk} used the Birkhoff theorem to derive the corresponding equations of motion (EOMs) describing the gravitational collapse of a thin shell. However, Zhang and Yi \cite{3} pointed out that there is a misunderstanding in the interpretation of the Birkhoff theorem and that the original result of 
\cite{2} 
is questionable. They argue that the correct metric, which is continuous with the location of an external observer, is determined by both the enclosed mass and the mass distribution outside. 
The statements in the work by Zhang and Yi were elaborated in \cite{Kim:2016osp} where it was demonstrated that the Birkhoff's theorem holds using a traditional  metric-based approach. They also show that it holds at the level of tetrad components. What they consider does not depend on the distribution of matter in the interior or exterior regions, or state of motion of the system, under the assumption that spherical symmetry holds. It would be unphysical though to rescale the interior vacuum region and keep the line-element outside of the spherical shell unchanged because it would led to discontinuity in the time coordinate and corresponding term in the metric element. An additional example of the gravitational field at a certain radius in a spherically-symmetric matter distribution which is outside of the radius has also been analyzed in \cite{Nandra:2013jga}.

Even though, this discussion on the different versions of Birkhoff's theorem is interesting from a theoretical perspective, it is irrelevant for our purposes. The reason for this irrelevance is that 
in the non-relativistic low curvature limit both versions of the metrics give same result for the stability of the shell in our construction. 
We decided to follow the 
formulation used by (\cite{2,Brevikk}), but in our case, (\cite{Kim:2016osp}) would have given the same results.

\subsection*{B $\quad$ 
Analytic expressions for temperature dependence}

The analytic expressions for the roots of the EOM (\ref{eq_scalarLowTemp}) are
\begin{tiny}
  \bea \label{eq_R0forTscalar}
R_1&=&\frac{1}{36 T \zeta (3)}
\left(
\sqrt[3]{216 \zeta (3) \left(\sqrt{\left(\text{C}-G_N
   m_S^2\right) \left(11664 \zeta (3)^2
   \left(\text{C}-G_N m_S^2\right)-\pi ^9\right)}-108
   \zeta (3) \left(\text{C}-G_N m_S^2\right)\right)+\pi
   ^9}\right.
   +\\ \nonumber
   &&
   \left.
   \frac{\pi ^6}{\sqrt[3]{216 \zeta (3)
   \left(\sqrt{\left(\text{C}-G_N m_S^2\right)
   \left(11664 \zeta (3)^2 \left(\text{C}-G_N
   m_S^2\right)-\pi ^9\right)}-108 \zeta (3)
   \left(\text{C}-G_N m_S^2\right)\right)+\pi ^9}}+\pi
   ^3\right)\\
   R_{2,3}&=&
   \frac{1}{72 T \zeta (3)}\left(\pm i \left(\sqrt{3}+i\right) \sqrt[3]{216 \zeta (3)
   \left(\sqrt{\left(\text{C}-G_N m_S^2\right)
   \left(11664 \zeta (3)^2 \left(\text{C}-G_N
   m_S^2\right)-\pi ^9\right)}-108 \zeta (3)
   \left(\text{C}-G_N m_S^2\right)\right)+\pi
   ^9}+\right.\\ \nonumber
   &&\left.
   \frac{\pm \left(-1-i \sqrt{3}\right) \pi ^6}{\sqrt[3]{216 \zeta (3)
   \left(\sqrt{\left(\text{C}-G_N m_S^2\right)
   \left(11664 \zeta (3)^2 \left(\text{C}-G_N
   m_S^2\right)-\pi ^9\right)}-108 \zeta (3)
   \left(\text{C}-G_N m_S^2\right)\right)+\pi ^9}}+2 \pi
   ^3\right)
  \eea
\end{tiny}
Interestingly, the temperature $T$ factors out of the solutions (\ref{eq_R0forTscalar}), thus the number of real roots only depends on the values of $(C,\, G_N, \,$ and $m_S)$. Further, $C$ has a definite value  $C=0.0461766$~\cite{Bordag:2009zz}. Thus, we can 
multiply both sides of (\ref{eq_R0forTscalar}) with $T$
to make it a dimensionless identity.
Now we can express all dimensionful quantities in Planck units, which leaves $m_S$ as the only parameter which determines the number of real roots.

\subsection*{C $\quad$ Inclusion of cosmological constant in the interior}
\textcolor{black}{
For the illustrative purposes, on the Figure 6 (\ref{lambdainfluence}), we show the effect of including cosmological constant in the interior of the mass shell. For this effect to be visible the cosmological constant needs to be very large. For the realistic values of the cosmological constant, these effects can be neglected.
\begin{figure}[ht!]
    \centering
    \includegraphics[width=7cm]{cconst1.pdf}  \includegraphics[width=7cm]{cconst1Log.pdf}            
    \caption{ 
    \textcolor{black}{
    The $G_N$ is fixed to 1, while other parameters are $C = 0.0028 , m_S = 0.034$, and values of  $\Lambda_i$ are taken to $ \Lambda_i= 0.000001, 0.00001, 0.0001, 0.001, 0.01$ for the blue, orange,  green, red and  purple graphs, respectively.
    }
    }
    \label{lambdainfluence}
\end{figure}
}

%
\subsection*{D $\quad$ Comment on the null energy condition}

We want to explore whether the null energy condition and its possible violation could be related to the existence of a stable configuration.
We implemented the interplay between the Casimir effect and Einsteins field equation at the level of a sum of effective forces rather than at the level of a full fledged theory combining gravitational and quantum concepts.
Thus, to be consistent with our approach, we also have contemplate the NEC from both sides of the problem separately. 
\begin{itemize}
    \item The gravitational side, is represented in terms of the Einstein Tensor $G_{\mu \nu}$.
    \item 
    The quantum side, is represented in terms of the corresponding averaged Casimir energy and pressure.
\end{itemize}
In the following subsections we calculate both types of NEC conditions and find that they are both fulfilled. 

\subsubsection*{Gravitational perspective}

A crucial aspect of the
gravitational system is that one distinguishes between inside and outside regions with respect to the shell.
Thus, we also discuss the NEC in both regions.
\begin{itemize}
\item 
We consider the ansatz of a radial vector $n^a=(n_1(t,r),n_2(t,r),0,0)$, which is null when $n_an_bg_{interior}^{ab}=0$ and we find 
\begin{align}\label{eq_NullVec}
n^a=n_1\left(1,\frac{1-\Lambda r^2/3}{\sqrt{R}}\sqrt{R-2 G_N m},0,0\right).
\end{align}
The null energy condition $T_{ab}n^an^b$ can be inferred from assuming that Einsteins field equations 
\be\label{eq_Einstein}
G_{\mu \nu}- \Lambda g_{\mu \nu}= 8 \pi G T_{\mu \nu}
\ee
hold locally inside the shell. The left hand side of these equations, can be calculated from the line element given in (\ref{eq_ds2int}).
Contracting (\ref{eq_Einstein}) with the null vectors (\ref{eq_NullVec})
we find
\begin{align}
T_{ab}n^an^b=0.
\end{align}
which, saturates the NEC trivially.
\item In the exterior we have Schwarzschild metric, or de-Sitter Schwarzschild metric
which have a vanishing stress-energy tensor, which implies
$T_{ab}n^an^b=0$.
\end{itemize}

\subsubsection*{Vacuum-matter perspective}

For a massless scalar field, we obtain from equation \eqref{ecasimir} the average net energy density to be
\begin{align}
    \rho = \frac{E}{V} = \frac{C}{2R}\frac{3}{4\pi R^3} = \frac{3C}{8\pi R^4}\,.
\end{align}
Similarly we obtain from equation \eqref{pcasimir} the average net pressure
\begin{align}
    p  = \frac{C}{8\pi R^4} \,.
\end{align}
Both quantities are calculated in flat space-time. Thus, the NEC in this context is simply given by the sum of
energy and pressure densities 
\begin{equation}
    \rho + p = \frac{C}{2\pi R^4}\,.
\end{equation}
Since both the constant $C$ and the radius $R$ are positive, we conclude that $\rho + p \geq 0$ $\forall R\geq0$. Differently, for a massive scalar field, using the fitting function \eqref{eq_fitREm}, we find an energy density
\begin{equation}
    \rho = \frac{E_m}{V}=\frac{3 (a+b m R)}{4 \pi  R^4 \left(c m R+d m^3 R^3+1\right)}\,,
\end{equation}
and the pressure $p=-(\partial_R E_m) /(4 \pi R^2)$ gives
\begin{align}
    p =\frac{a + 2 a c m R + b c m^{2} R^{2} + 4 a d m^{3} R^{3} + 3 b d m^{4} R^{4}}
{4 \pi\, R^{4}\,\left(1 + c m R + d m^{3} R^{3}\right)^{2}}.
\end{align}
The sum of the energy and pressure density then gives
\be
    \rho + p =
    \frac{ b\,mR\left(3 + 4cmR + 6dm^{3}R^{3}\right) + a\left(4 + 5cmR + 7dm^{3}R^{3}\right)}
{4\pi\,R^{4}\left(1 + cmR + dm^{3}R^{3}\right)^{2}}
   \,.
\ee
For the numerical values of the constants given in (\ref{eq_fitREm}), this quantity is positive (even for $b=0$). Thus, for these values the stability condition is not related to a NEC violation.

\end{document}